\documentclass[aps,twocolumn,showpacs,preprintnumbers]{revtex4-1}

\usepackage[utf8]{inputenc}
\usepackage[T2A]{fontenc}

\usepackage{graphicx}  
\usepackage{subfigure}
\usepackage{multirow}

\usepackage{fancyhdr}
\usepackage{longtable}
\usepackage{parskip}
\usepackage[T1]{fontenc}
\usepackage{dcolumn}   

\usepackage{bm}        
\usepackage{amsfonts}  
\usepackage{amsmath}   
\usepackage{amssymb}   

\newcommand{\im}[1]{\mathrm{Im}\left(#1\right)}

\begin{document}

\title{Phonon-Limited Mobility in H/F-functionalized Nanotubes with 1D $\pi$-chains}

\author{V.L. Katkov}

\affiliation {\it  Bogoliubov Laboratory of Theoretical Physics,  Joint Institute for Nuclear Research, Dubna,  Russia}

\author{V.A. Osipov}

\affiliation {\it Bogoliubov Laboratory of Theoretical Physics,  Joint Institute for Nuclear Research, Dubna, Russia}


\begin{abstract}  
Electron mobility due to electron-phonon interaction is investigated for fully fluorinated/hydrogenated zig-zag carbon nanotubes containing one-dimensional alternating chains of carbon atoms with $\pi$-bonds. The behavior of mobility associated with changes in the tube diameter, coating type (F/H) and temperature is revealed. In particular, it is shown that the dependence of mobility on the diameter in such tubes is periodic with the chirality index, which is associated with the absence of scattering on TA phonons in the tubes with an even number of conducting chains due to mirror symmetry. The obtained small values of phonon-limited mobility indicate that tubes with one-dimensional conducting chains are more promising for use as gas sensors than as elements of electronic devices. Calculations are performed within the self-energy relaxation time approximation (SERTA) using the non-orthogonal tight-binding approach.
\end{abstract}

\pacs{47.15.-x}

\maketitle 

\section{Introduction}


Carbon nanotubes (CNTs) are used in electrical applications as both switching elements and conductive connectors due to their unique electrical properties, including high electrical conductivity and tunable resistance. In switches, CNTs can be electrostatically controlled to make or break an electrical circuit by moving them into contact with different electrodes. For connectors, CNTs can be dispersed in polymers to create conductive films and fibers that replace heavier traditional materials like copper~\cite{Sanial2017}. This is supported by the expectation that with a small number of defects, sufficiently large mean free paths should ensure the preservation of the ballistic type of transport for long tubes. At the same time, the key value characterizing the efficiency of a particular material for electronic, optoelectronic and thermoelectric applications is the mobility of charge carriers.
   
Among various factors that influence the mobility of carriers in CNTs electron-phonon interactions are central in understanding their transport properties. There have been many theoretical studies on the calculation of carrier scattering rates and mobilities in CNTs using the Boltzmann transport equation (often solved with Monte Carlo methods or iterative techniques) and first-principles calculations (see, e.g., a recent review~\cite{Claes2025}). Notice that first-principles analysis is a complex and resource-intensive task, which explains the fact that today the range of materials where phonon-limited mobility is calculated by such approaches is restricted.

The object of our study is carbon nanotubes of the zig-zag type, chemically modified by the addition of hydrogen or fluorine (both elements are monovalent, therefore they form similar structures \cite{Sorokin2010}). It has been shown experimentally previously that the maximum concentration of F as a modifying impurity is 50~\%. \cite{Mickelson1999}. Theoretical calculations confirm the stability of such maximally doped tubes with both F and H~\cite{Seifert2000, Kudin2001, Bauschlicher2001}.
Moreover, the most energetically preferred configurations are those with the formation of $\pi$-chains of carbon atoms. In particular, they are more favorable than configurations with $\pi$-$\pi$ dimers, located both regularly and randomly~\cite{Bauschlicher2001}.

There are two configurations of $\pi$-chains: “helical”, when $\pi$-chains are arranged in the form of a spiral, and “chain”, when $\pi$-chains are extended along the axis of the tube (see Fig. 1).
Previously, we calculated the temperature dependence of the bandgap for such CNTs and discovered fundamentally different behavior of this value for “chain” and “helical” nanotubes~\cite{Katkov2023}. 
\begin{figure}
\includegraphics[width=2.5in]{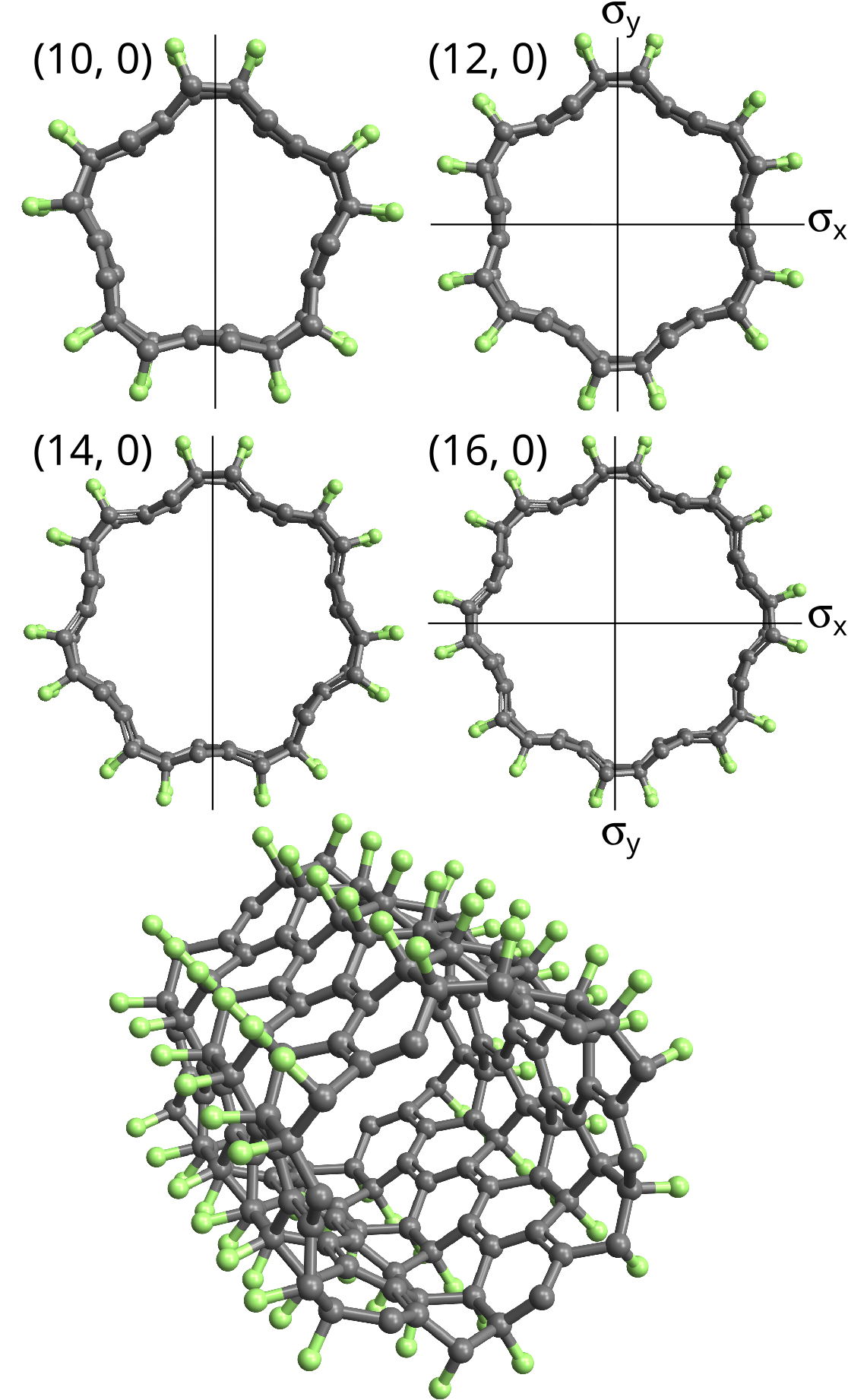}
\caption{\label{tubes} 
Functionalized by F/H carbon nanotubes in a “chain” configuration. It is evident that with an increase in the chirality indices, the number of one-dimensional $\pi$-chains increases by one from 5 for (10, 0) to 8 for (16,0) tube. }
\end{figure}

In this paper, we restrict our consideration to only zig-zag type CNTs with the “chain” configuration. This is motivated by the fact that their bandgap varies in the range of 0.3-0.8 eV, which is more attractive for use in various semiconductor devices. For comparison, in the “helical” structures the bandgap exceeds 1 eV. Moreover, in contrast to the “helical”,  for “chain”  configuration the diameter of the nanotube has a significant effect on the bandgap~\cite{Kudin2001, Katkov2023}, which implies the possibility of fine-tuning the gap.

It is interesting that electron transport in such tubes occurs along one-dimensional \textit{cis}-polyacetylene “channels”. In the case of ballistic regime, such a study was carried out in \cite{Ranjan2006} using both the DFTB method \cite{Porezag1995} and the nonequilibrium Green's function method \cite{Katkov2017}. However, this consideration does not answer the question of what effect the “channels” will have in the case of diffusion transport. Moreover, it is not entirely clear at what length of tubes the ballistic regime will be maintained. Our goal is to study the transport properties of “chain” CNTs caused by the electron-phonon interaction. In particular, we calculate the mobility and the mean free path of electrons, which in turn allows us to estimate the maximum length of tubes at which ballistic transport is possible. Zig-zag type CNTs of various diameters, with two modifying impurities (H or F) and at different temperatures are considered.


\section{Model}

Electron mobility is defined as
\begin{equation}\label{eq.1}
    \mu = \frac{\sigma}{e n},
\end{equation}
where $\sigma$ is the conductivity, and $n$ is an electron concentration. Normalized per unit area of the tube surface, $n$ takes the form
\begin{equation}\label{eq.2}
    n = \frac{1}{a_z L}\int \limits_{E_F}^\infty d\varepsilon \rho(\varepsilon)f(\varepsilon),  
\end{equation}
with $\rho(\varepsilon) = 4 a_z/h  \sum\limits_n v_n^{-1}(\varepsilon) \theta(\varepsilon- E_C^n) $ being the density of states per unit cell, $E_C^n$ the bottom of the conduction band of the $n$-th branch, $E_F$ the Fermi level, $v_n(\varepsilon)= \partial E_n/\partial p$ the group velocity of the $n$-th branch, $a_z$ the length of the translation vector of the unit cell, and $L$ the tube perimeter.

In the approach associated with solving the Boltzmann transport equation (BTE), the conductivity of CNTs, normalized to the length of the tube perimeter, is as follows (the $\sigma$ dimension is Siemens)~\cite{ Fang2008, Zhao2009, Ponce2020, Unsal2025}
\begin{eqnarray}\label{eq.3}
    \sigma &=& \frac{2 e^2}{a_z L} \sum_n\int\limits_{-\pi/a_z}^{\pi/a_z} \frac{dk }{\Omega_{BZ}}\left[-\frac{\partial f_{nk}}  
    {\partial\varepsilon_{nk}}\right] v_{nk}^2 \tau_{nk} \\ \nonumber 
    &=& \frac{4 e^2}{h L} \sum_n\int\limits_{E_C^n}^{\infty} d \varepsilon \left[-\frac{\partial f(\varepsilon)}  {\partial\varepsilon}\right]v_{n}(\varepsilon)\tau_{n}(\varepsilon).
\end{eqnarray}
Here  $\tau_n(\varepsilon)$ is the relaxation time  of the $n$-th branch and $\Omega_{BZ} = 2\pi/a_z$.  The main problem is the calculation of $\tau(\varepsilon)$ or, equivalently, the mean free path $\lambda(\varepsilon, T) = v(\varepsilon)\tau(\varepsilon, T)$. 

There are several approximations for solving the diffusion transport problem (for more details, see \cite{Claes2025, Ponce2020}). The most accurate is the iterative solution of BTE.
Less accurate, but also less expensive, is the calculation of $\tau$ using SERTA (self-energy relaxation-time approximation). The relaxation time calculated in this approximation is uniquely related to the imaginary part of the Green's function and characterizes the broadening of the electron energy level due to the finite lifetime: $\im{\Sigma} = -\hbar/2\tau$. The roughest and fastest can be considered the calculation of $\tau$ through the deformation potential, which is estimated by the shift of the energy bands when the structure is stretched as is done, for example, in \cite{Long2011, Xi2012}. However, the latter method is very inaccurate and the discrepancy with the mobility calculated in SERTA can reach three orders of magnitude
 \cite{Unsal2025}. At the same time, even in extreme cases, the error in calculating mobility in SERTA does not exceed 70\% compared to more accurate and expensive calculations~\cite{Claes2025}.

The SERTA approach has been implemented in several software packages based on the density functional theory (DFT)~\cite{Claes2025}. However, the main drawback of using DFT is that the computational resources required to calculate $\tau$ increase disproportionately with the number of atoms in the unit cell. In our case of functionalized CNTs, this number is quite large (about $10^2$ atoms).

It should be noted that the electron-phonon interaction constant and the electron relaxation time  can be calculated not only using the DFT method, but also within the tight-binding approximation. 
This idea was first used in \cite{Barisich1970} for a simple cubic lattice (the so-called BLF or BLF-SSH model). The method was then generalized to non-orthogonal basis \cite{Varma1979}. Recently, this approach was adapted to work with the DFTB+ code  (a popular package implementing the non-orthogonal tight-binding method)~\cite{Hourahine2020} within the DFTBephy python package \cite{Croy2023}. In particular, DFTBephy allows one to calculate the relaxation time in the SERTA approximation. It turned out that its results agree well with first-principles calculations for various materials. At the same time, the calculations become orders of magnitude less resource-intensive. Moreover, they can be implemented on modern desktop computers, which makes this approach the most suitable tool for solving the problem.

Briefly, the DFTBephy approach is structured as follows. Using the non-orthogonal tight-binding method implemented in the DFTB+ as well as the Phonopy package~\cite{phonopy2015, phonopy2023, phonopy20232}, the phonon characteristics are calculated on a supercell of sufficient size (all coordination spheres for which the used tight-binding parameters are calculated must be included). Then, for this supercell, the atoms are successively shifted by a small amount in three directions and the derivatives with respect to $x$, $y$ and $z$ of the hopping and overlapping parameters of the atomic orbitals are calculated (DFTB+ allows saving the real-space Hamiltonians $\textbf{H}$ and the overlap matrices $\textbf{S}$ as text files). The next step is to expand the obtained matrix elements into a Fourier series with respect to the positions of the atoms. Knowledge of these quantities, the eigenvectors and eigenvalues of the Schrödinger equation, the phonon frequency spectrum, and the corresponding polarization vectors will be sufficient to calculate the electron-phonon coupling constants and the relaxation time in the SERTA approximation. A detailed description of the approach is given in \cite{Croy2023, Unsal2025}.

Note that the tight-binding method was previously used to calculate the phonon-limited mobility in pure carbon nanotubes of various diameters and graphene~\cite{Li2015}. However, this method was not universal. To implement this approach, the authors had to use both the hopping parameters and their dependence on the distance between atoms for the electron subsystem as external data, and also independently determine the force constants to describe the phonon subsystem. Neither of these are known in the general case, but can be obtained using additional DFT calculations. At the same time, in DFTB+ both the hopping parameters responsible for electron transport and the repulsion parameters responsible for elastic properties are already known for the main pairs of chemical elements (previously calculated based on DFT calculations) and are presented in several ready-made sets. This makes the tight-binding method with a non-orthogonal basis more universal and suitable for calculating mobility not only in graphene and carbon nanotubes, but also in various functionalized materials.

In our calculations we use DFTB+ with self-consistent charge calculation (SCC). The parameterization method for the hopping integrals was chosen as GFN1-xTB \cite{Grimme2017} (tests showed that the results do not change significantly when choosing the matsci-0-3 or pbc-0-3 parameterization, but GFN1-xTB gives phonon spectra for pure carbon nanotubes that are closer to the DFT results~\cite{Dubay2003}). Phonon characteristics were calculated using the Phonopy code, relaxation time using DFTBephy adapted to the SCC and GFN1-xTB methods. Supercells were chosen large enough to avoid negative frequencies in the spectrum (7 or 9 unit cells). The calculation was carried out over the entire Brillouin zone of phonons, similar to the work of \cite{Li2015}. Temperatures from 10 to 300 K and electron concentrations up to $2.5\times 10^{13}$~cm$^{-2}$ were considered. With this choice of parameters, the chemical potential was always in the vicinity of $E_C$. To test the approach used, we recalculated several points from Fig. 2(a) in~\cite{Li2015} and reproduced both the mobility value of pure CNTs and its qualitative behavior with diameter.

\section{Results}

Calculations were performed for functionalized with F/H CNTs of the “chain” type shown in Fig.~\ref{tubes}. In all cases, the tubes are semiconducting due to the fact that the $\pi$-chains responsible for the electron branches closest to the Fermi level are dimerized. Moreover, with an increase in the tube diameter, both the degree of dimerization (the ratio of the lengths of two consecutive bonds) and the value of the energy gap increase and both values tend to saturate (see Table~\ref{t1}).

\begin{table}
	\caption{Energy gap for fully fluorinated/hydrogenated carbon nanotubes of different chirality (in eV).}\label{t1}
	\begin{tabular}{c| l| l| l |l }
		\hline \hline
		  ~~~~~  & (10,0) & (12,0)  & (14,0)& (16,0)    \\
				
		\hline
		H &  0.45&  0.61&  0.77 & 0.85  \\
		\hline
            F & 0.33  & 0.47 &  0.61 & 0.68  \\
		\hline
		\hline \hline
		
	\end{tabular}
\end{table}

\begin{figure}
\includegraphics[width=2.5in]{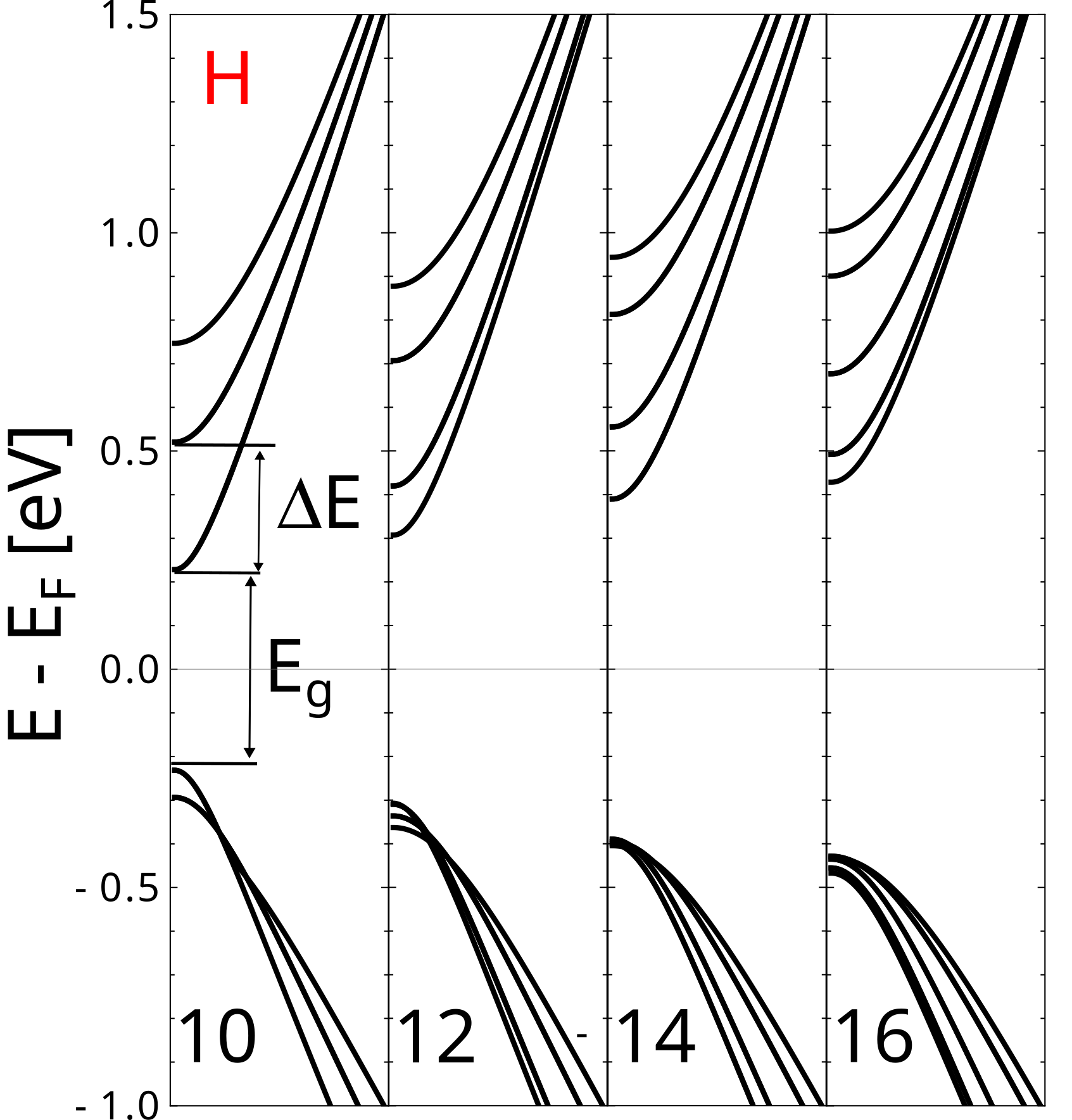}
\caption{\label{bands} 
Band structure near the Fermi level for H-functionalized “chain” carbon nanotubes of different diameters. All branches of $\pi$-chains in both the conduction and valence bands are shown. Values of n are indicated along the $x$ axis. For tubes with n=10 and 14, the lower level is doubly degenerate. F-functionalized CNTs have a similar band structure (not shown). $\Delta E$ is given by Eq. (\ref{eq.4}).}
\end{figure}

In the conduction band, the $\pi$-chains correspond to the branches closest to the Fermi level (see Fig.~\ref{bands}). They are a set of parabolic curves (channels) whose minimum is at $k = 0$. The number of such channels $N$ is equal to the number of $\pi$-chains in the CNT (for our tubes $N=$ 5, 6, 7, and 8, respectively). The position of the minima for all CNTs quite accurately obeys a relationship similar to the spectrum of atoms in a periodic monatomic chain~(Born-von Karman boundary conditions): 
\begin{equation}\label{eq.4}
    \Delta E_{F,H}(m) =  t_{F,H} \cos\left(\frac{2\pi m}{N}\right)-E_{min},
\end{equation}
where $m$ takes integer values from 0 to $N-1$, and $E_{min}$ is the minimum value from the set $\{t_{F,H}\cos\left(2\pi m/N\right)\}$. Moreover, for all structures with fluorine, $t_F$ can be estimated as 0.14~eV, whereas  for all structures with hydrogen, $t_H\approx 0.25$~eV. This means that the $sp^3$ rows of H atoms bonded to carbon isolate the $\pi$ chains less strongly than the similar rows of F atoms.

As follows from Eq. (4), for even $N$ the lowest energy level will not be degenerate since in the set of $m$ there will always be one value at which the cosine turns to $-1$.
For odd $N$ the cosine takes the smallest value at two points $\pi \pm \pi/N$, so the lower zone will always be doubly degenerate.
As our calculations have shown, in a wide range of temperatures (up to 300 K) and concentrations (up to $2.5\times 10^{13}$ cm$^{-2}$) the conductivity is determined by the lowest branch of the spectrum.

One of the quantities that determines the mobility~(\ref{eq.1}) is the group velocity $v_{n}$, which is manifested both in concentration~(\ref{eq.2}) and in conductivity~(\ref{eq.3}). Moreover, within the deformation potential approximation it influences the value of $\tau(\varepsilon)$: $\tau(\varepsilon)\propto 1/\rho(\varepsilon)\propto v(\varepsilon)$. 
\begin{figure}
\includegraphics[width=2.5in]{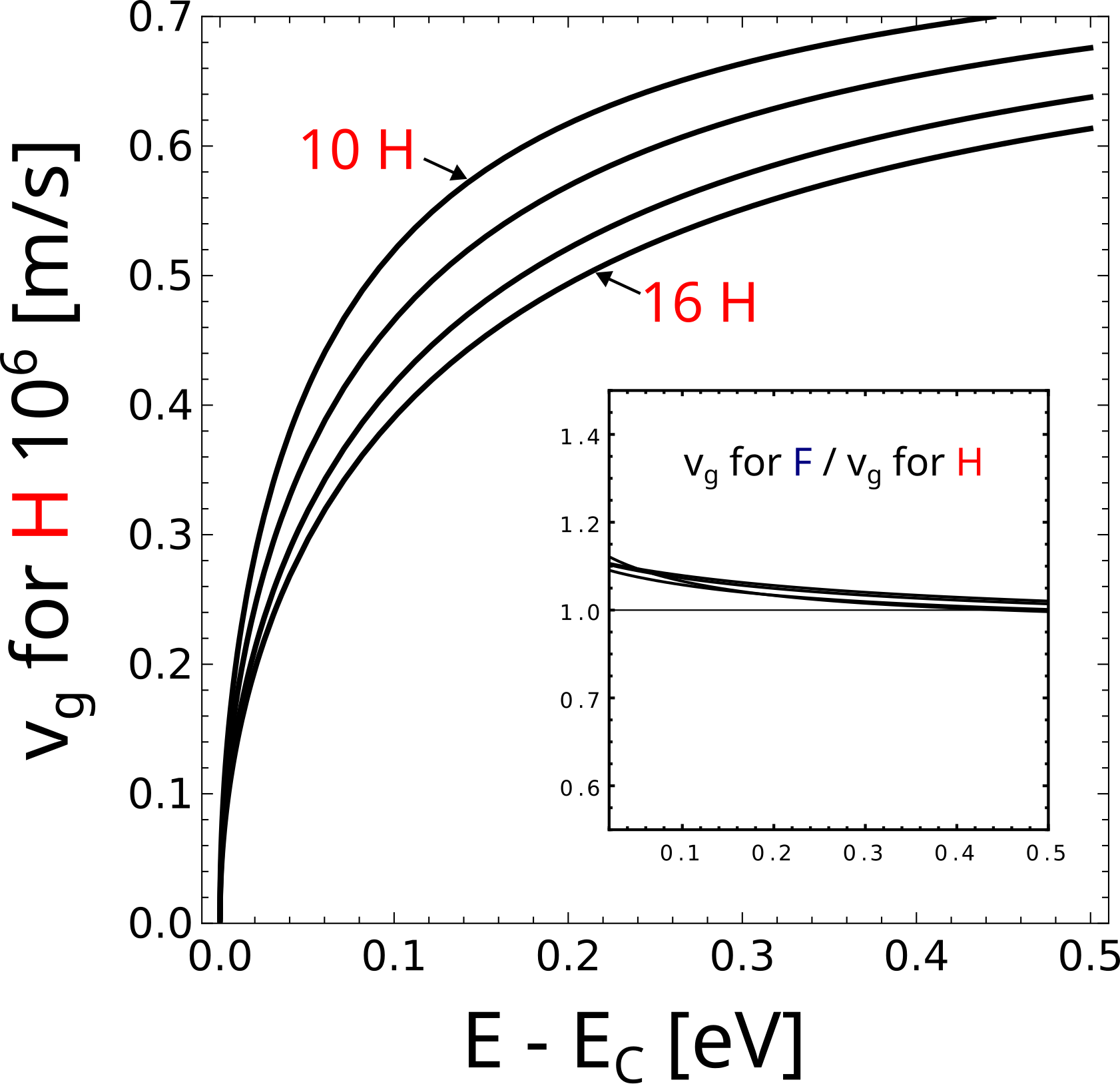}
\caption{\label{vel} 
Group velocity for the first branch of hydrogenated CNTs as a function of energy measured from the bottom of the conduction band. The curves run sequentially from top to bottom from the tube with the smallest diameter (10, 0) to the tube with the largest diameter (16, 0). The inset shows the ratio $v_{1,F}(\varepsilon)/v_{1,H}(\varepsilon)$.
}
\end{figure}
Fig.~\ref{vel} shows $v_{1}$ as a function of energy for hydrogenated CNTs for the lowest branch of the conduction band (the energy is measured from $E_C$ ). A noticeable decrease in velocity is seen with increasing tube diameter. That is, from considerations related only to group velocity, mobility should decrease with increasing tube diameter. The inset shows the ratio of the group velocities of fluorinated and hydrogenated CNTs $v_{1,F}(\varepsilon)/v_{1,H}(\varepsilon)$ as a function of energy. 
Note that although the velocities are close, there is a difference of approximately 10 \% in the most significant region near $E_C$.

\begin{figure}
\includegraphics[width=2.5in]{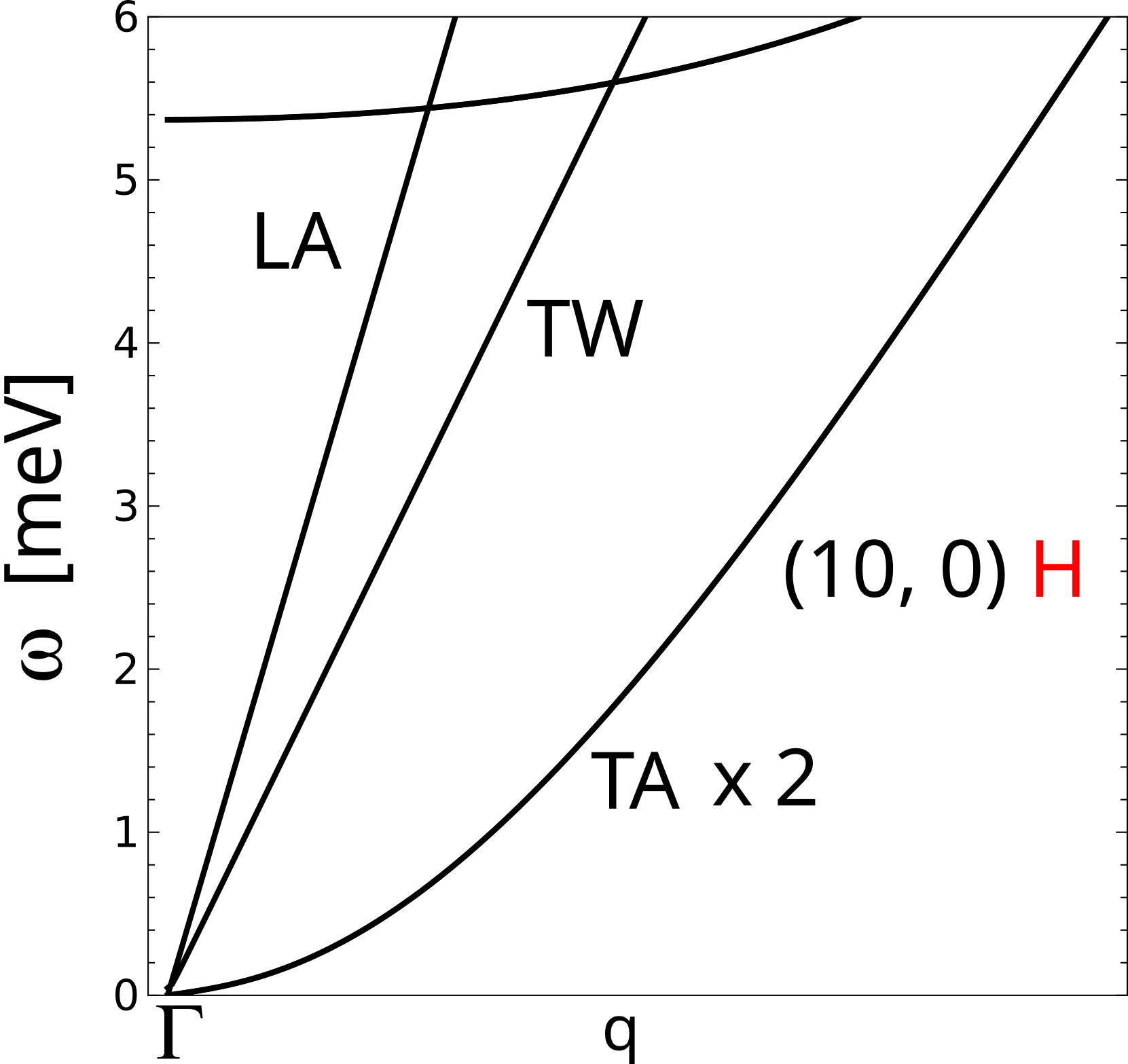}
\caption{\label{ph} 
Spectrum of acoustic phonons near the $\Gamma$ point for hydrogenated (10, 0) CNT. From zero there emerge a doubly degenerate quadratic branch TA, as well as linear branches LA and TW.}
\end{figure}

The main contribution to the scattering rate $\tau^{-1}(\varepsilon)$ at low $\varepsilon$ is made by the lowest-energy branches of the \textit{phonon} spectrum of CNTs, intersecting at the point $\omega = 0$ (this is due to the influence of the singularity of the density of \textit{electron} states at $E_C$). In this region, the structure of the phonon branches $\omega_n(q)$ of all functionalized tubes is similar to each other and is similar to that of pure carbon nanotubes \cite{Saito1998}.
In particular, four acoustic branches intersect at the $\Gamma$ point. As an example, Fig.~\ref{ph} shows the phonon spectrum for hydrogenated (10,0) CNT. The spectra of fluorinated tubes  are similar, but the branches are located slightly lower. 
The first two branches are identical (merge into one) and have a quadratic dependence on $q$, which is typical for low-dimensional structures. These are transverse acoustic (TA) branches.  The next is the so-called twisting (TW) mode with linear behavior, and then the longitudinal acoustic (LA) mode.

At small $\varepsilon$, we found that for odd $N$, the main contribution to $\tau^{-1}(\varepsilon)$ comes from the TA branches, while for even $N$, their contribution is negligible. 
This follows from the fact that in nanotubes with even $N$ the $\pi$-chains are arranged mirror-symmetrically relative to the $\sigma_x$ and $\sigma_y$ planes parallel to the tube axis (see Fig.~\ref{ph}). Due to the orthogonality of the polarization vectors of the TA modes and their oddness relative to the mirror reflections in these planes, a strong suppression of the electron-phonon coupling occurs. For odd $N$ the symmetry is lower and such compensation does not occur.
This situation is analogous to the suppression of electron-phonon scattering for the out-of-plane ZA mode in graphene or in nanoribbons with mirror symmetry~\cite{Chu2014}.
In all the cases considered, LA and TW modes make a comparable contribution to $\tau^{-1}(\varepsilon)$. 

\begin{figure}
\includegraphics[width=3.0in]{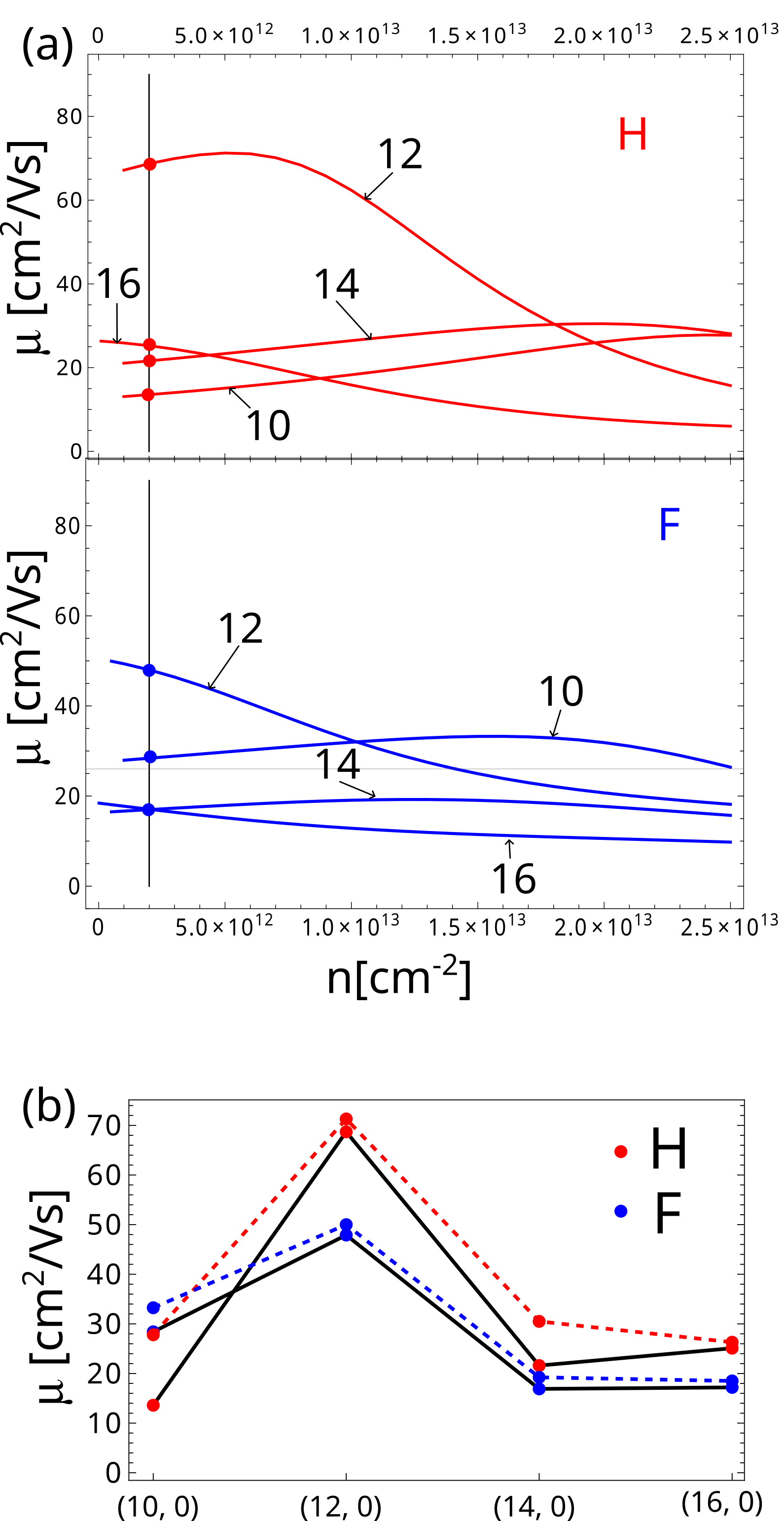}
\caption{\label{mu} (a) Mobility as a function of concentration for all tubes considered. (b) Mobility as a function of diameter: solid lines correspond to a fixed concentration $n = 2\times 10^{12}$~cm$^{-2}$ (intersection points with the vertical curve in the upper plot), dashed lines correspond to the peak (maximum) mobility; $T = 300$ K.
}
\end{figure}

The mobility as a function of concentration is shown in Fig.~\ref{mu}~(a). The concentration $n = 2\times 10^{12}$~cm$^{-2}$ is indicated by the vertical line. All dependences are typical for pure carbon nanotubes (see, e.g., Fig. 2 in~\cite{Zhao2009}). The suppression of $\mu$ with increasing concentration is due to the inclusion of next electron subbands in the scattering process. A noticeable difference in the behavior of CNTs with even and odd $N$ is due to the different position of the next subband (see Fig.~\ref{bands} and Eq.~(\ref{eq.4})).

The electron mobility for the concentration $n = 2\times  10^{12}$~cm$^{-2}$ and the peak  mobility at  the temperature of 300 K is shown in Fig.~\ref{mu}~(b). The calculations were carried out using Eqs. (\ref{eq.1})-(\ref{eq.3}) for all $\pi$-branches.
The results for $H$ and $F$ tubes do not differ significantly, and in all cases the value of $\mu$ turns out to be small. 

\begin{figure}
\includegraphics[width=2.5in]{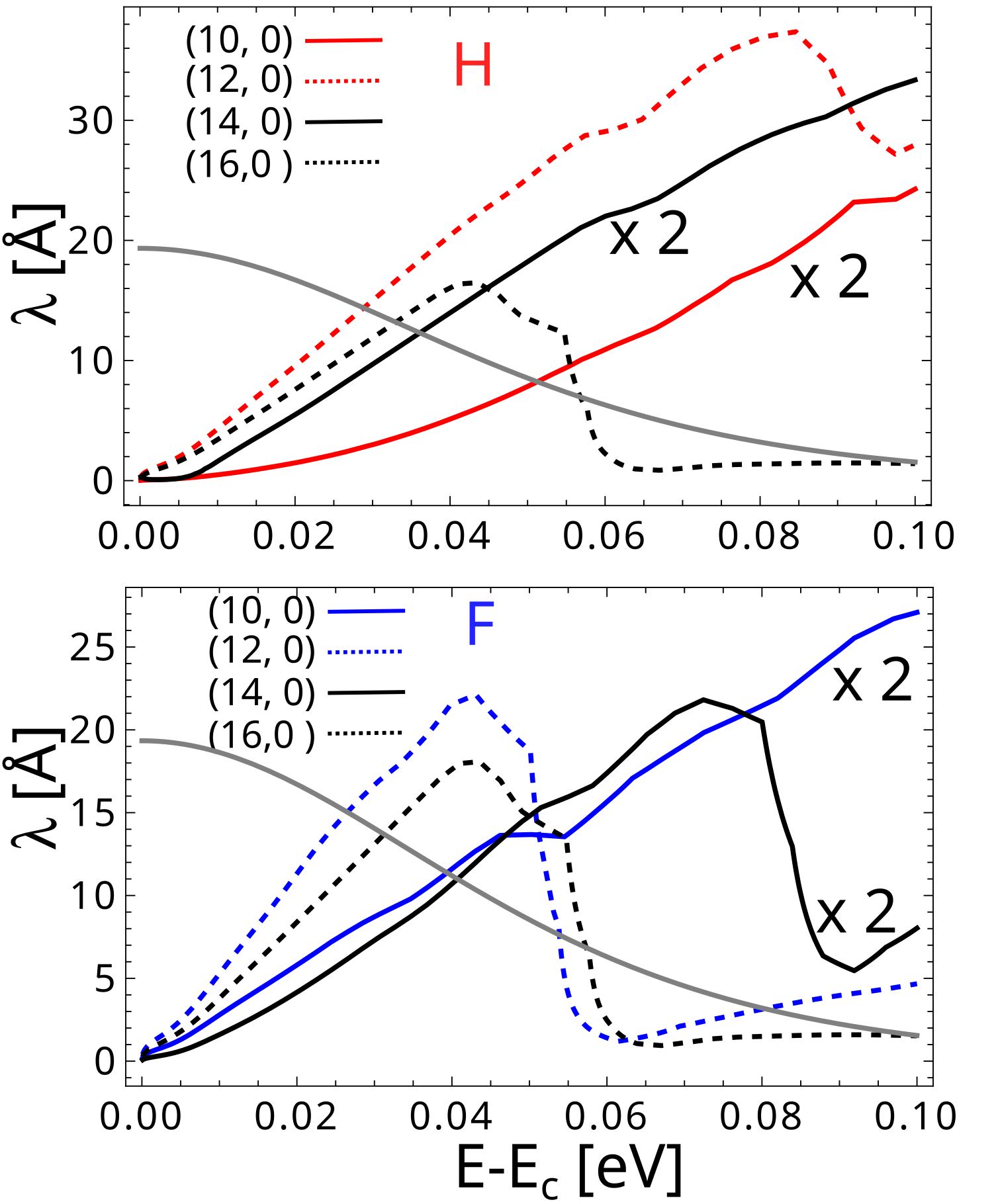}
\caption{\label{lenth} The mean free path $\lambda(\varepsilon) = v(\varepsilon)\tau(\varepsilon)$ for the lower branches of all the considered tubes at $n =2\times 10^{12}$~cm$^{-2}$ and $T = 300$ K. The solid colored curves marked by $\times 2$ are the result of the sum of two contributions of the doubly degenerated lower branch: $v_1(\varepsilon)\tau_1(\varepsilon)+ v_2(\varepsilon)\tau_2(\varepsilon)$.
The energy is counted from the bottom of the conduction band $E_C$. The gray curve shows the “window” for electrons that determine mobility, i.e. $-\partial f(\varepsilon)/\partial \varepsilon$ at $T = 300$~K and chemical potential $\mu = E_C$. 
}
\end{figure}

As can be seen, the monotonic behavior of the group velocity does not lead to monotonic behavior of the mobility when moving from (10,0) to (16,0). The mobility behaves periodically, for which the relaxation time is responsible.
To illustrate this, a plot of $\lambda(\varepsilon)$ for all tubes is given (see Fig.\ref{lenth}). If the lower branch is not degenerate, then the value $v_1(\varepsilon)\tau_1(\varepsilon)$ is given; if it is doubly degenerate, then the sum of equal contributions from each branch is given: $v_1(\varepsilon)\tau_1(\varepsilon) + v_2(\varepsilon)\tau_2(\varepsilon) = 2 v_1(\varepsilon)\tau_1(\varepsilon).
$
It is these products that make the decisive contribution to the integral (\ref{eq.3}). Also shown is the curve $-\partial f(\varepsilon)/\partial \varepsilon$ with the chemical potential located exactly at the level of $E_C$ at $T=300$ K. It defines the energy “window” of electrons participating in transport. The half-width of this function is $4 k_BT\approx0.1$~eV.

The mean free path $\lambda(\varepsilon, T)$ for the lower branch is always of the order of 10 \r{A}, which makes it difficult to observe ballistic transport in these structures at room temperature. Note that such small values are quite expected for structures in which the current is forced to flow along a single-atom chain and has no opportunity to “go around” the scattering center.

It should also be noted that in the deformation potential approximation one has  
$\lambda(\varepsilon)\propto v(\varepsilon)/\rho(\varepsilon)\propto \varepsilon$, which is roughly true for all the tubes we have considered. The sharp drop in $\lambda(\varepsilon)$ is due to the inclusion of a higher subband (branch) in the scattering process. Similar behavior occurs in pure carbon nanotubes~\cite{Zhao2009}.

For an odd number of chains $N$ the lower branch will be doubly degenerate, which would seem to lead to higher values of both conductivity and mobility. Actually, the situation is the opposite and is responsible for this relaxation time. The point is that the scattering rates $\tau_n^{-1}$ of electrons in a degenerate branch is noticeably higher because an additional scattering channel appears.
Indeed the scattering rate for branch $n$ is the sum of the partial scattering rates into itself and all other branches, according to “Matthiessen's rule”~\cite{Ponce2020,Unsal2025}:
\begin{equation}\label{eq.5}
    \tau_{n}^{-1} = \sum\limits_m \int\frac{d q}{\Omega_{BZ}}\tau^{-1}_{nm}(k , k+q).
\end{equation}
In the case of a non-degenerate branch, the contribution of $\tau_{11}$ will be the main one, since transitions to higher branches are suppressed for energy reasons. For a doubly degenerate branch, there will be two comparable contributions $\tau_{11}$ and $\tau_{12}$ due to coinciding branch. In this case there are two transport channels, and one would expect the scattering efficiency to be at least twice as high as for a single channel in the case of even $N$. However, in reality the efficiency is much higher, since, unlike in the case of even $N$, TA phonons are involved in the electron scattering.
 
 With increasing tube diameter, this effect decreases due to the fact that the number of branches participating in scattering becomes larger, which becomes the determining factor for tubes of large diameter. In particular, according to Eq.(\ref{eq.4}), with increasing $N$ the second subband is located closer to the first and the area of the “tooth”, which gives the main contribution to the integral (\ref{eq.3}), decreases (see  Fig.~\ref{lenth}).  Thus, periodicity in the behavior of mobility with increasing diameter appears for tubes of small diameter, but disappears with its increase.


\begin{figure}
\includegraphics[width=3.0in]{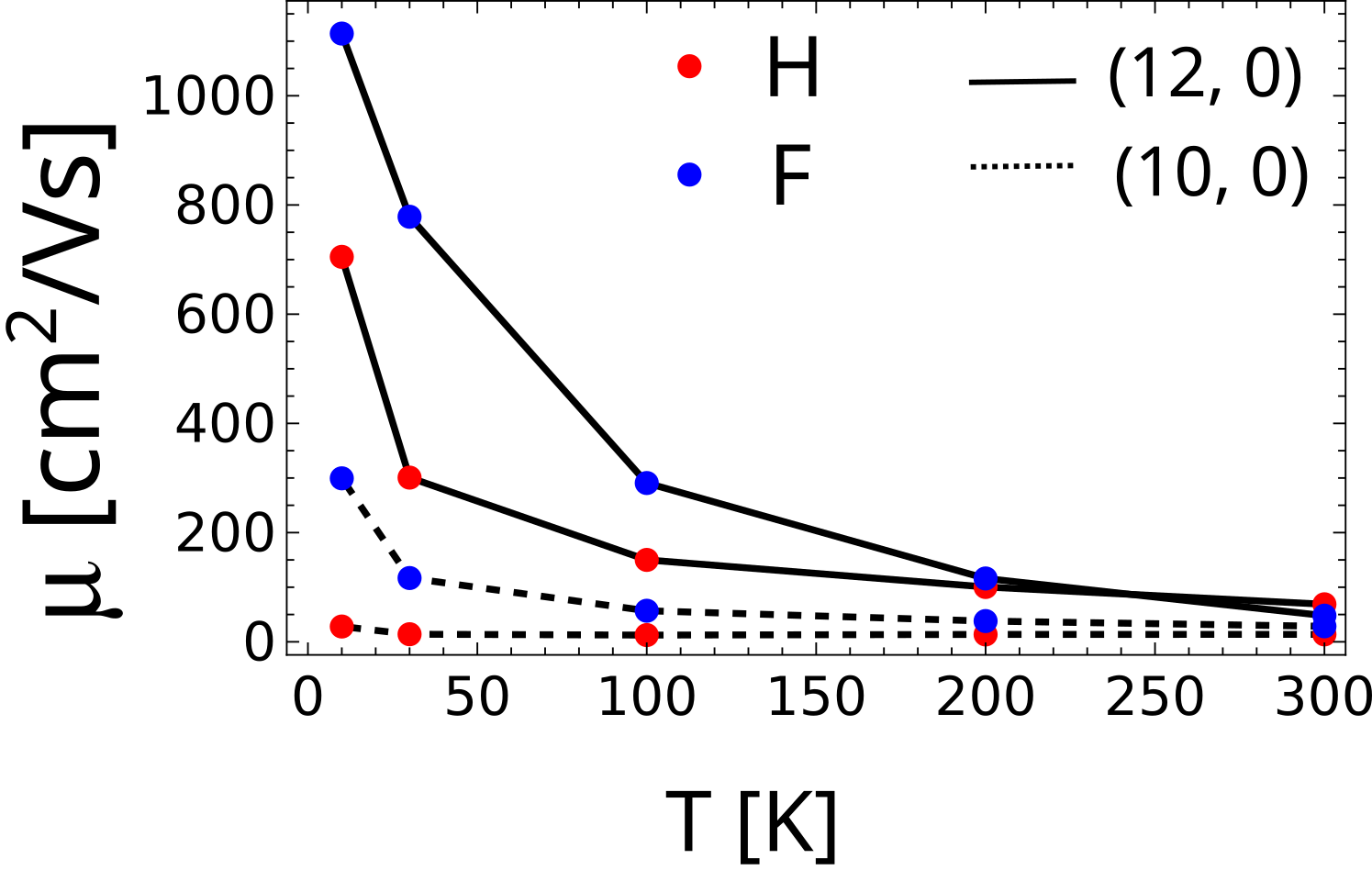}
\caption{\label{tdep} Mobility as a function of temperature at $n = 10^{12}$ cm$^{-1}$
}
\end{figure}

\begin{figure}
\includegraphics[width=3.0in]{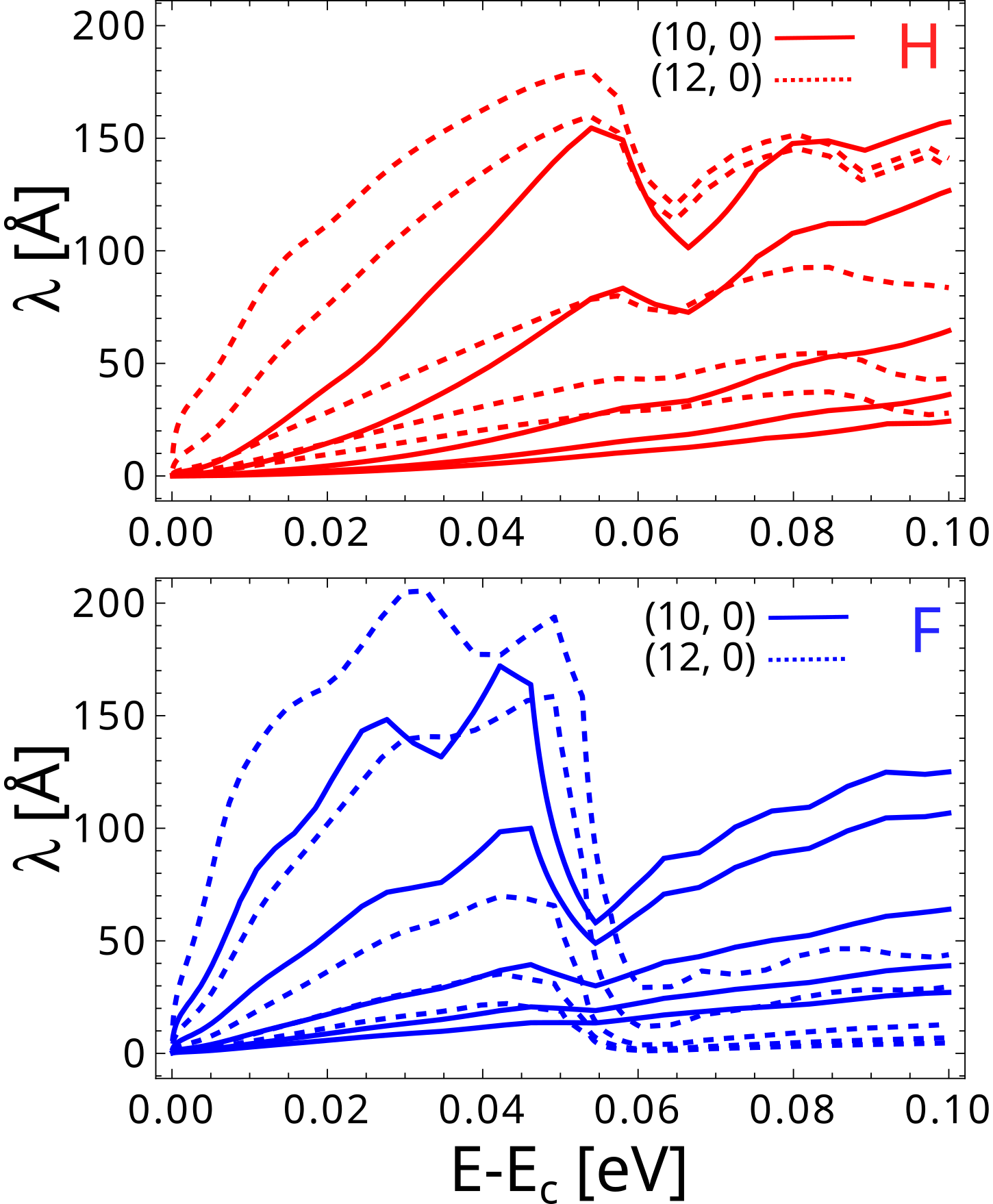}
\caption{\label{t-lambda} The mean free path as a function of energy at T = 300, 200, 100, 30, and 10 K (the corresponding curves are located from bottom to top) for tubes (10.0) and (12.0) with H and F. With decreasing temperature, $\lambda$ increases monotonically.
}
\end{figure}

The observed difference in the behavior of the scattering rate significantly affects the temperature dependence of the mobility in tubes with even and odd $N$. As shown in Fig.~\ref{tdep}, in tubes with a non-degenerate channel the mobility increases with decreasing temperature (approximately as $1/T$), which is typical for pure semiconducting CNTs~\cite{Perebeinos2005}. On the contrary, for tubes with a degenerate channel, the temperature dependence of mobility is weakly expressed, which can be explained by a more dynamic increase in the mean free path $\lambda$ with decreasing temperature for single-channel tubes, as shown in Fig.~\ref{t-lambda}. It should be noted that, although the expression for $\tau^{-1}$ in the SERTA approximation includes the value of the chemical potential, calculations in all the cases considered show that its influence is negligible.

An important conclusion is that, as can be seen from Fig.~\ref{t-lambda}, ballistic transport in the tubes under consideration can be detected only at extremely low temperatures and sufficiently high electron densities. Note that the suppression of $\lambda(\varepsilon)$ on all the curves shown in Fig.~\ref{t-lambda} is due to the inclusion of new optical branches in the scattering. As the temperature decreases, their influence becomes more noticeable. The exception here is the case of the fluorinated (12,0) tube (blue dotted line), where the suppression at an energy of 0.06 eV is due to the inclusion of the next subband.

\section{Summary}

We calculated the temperature-limited mobility and mean free path in carbon nanotubes with one-dimensional $\pi$-chains of different diameters functionalized with F or H (in both cases with a content of 50\%). The main results are as follows:
(i) the mobility in such tubes is found to be very small, which is a consequence of the one-dimensionality of the conducting channels, (ii) electron ballistic transport in these tubes is impossible at room temperatures due to the too small mean free path, (iii) unlike pure carbon nanotubes, the group velocity decreases with increasing tube diameter, (iv) the behavior of diffuse transport for tubes doped with either H or F is very similar, (v) the dependence of mobility on the tube diameter is not monotonic, as in pure CNTs, but “periodic” with a change in the chirality index n, (vi) in carbon nanotubes with odd $N$ and a degenerate lower branch, the dependence of mobility on temperature is expressed much weaker than for tubes with even $N$.

The low mobility of functionalized CNTs with $\pi$-chains over a wide temperature range makes them unpromising in applications related to their use in electronic devices, for example, as a transistor channel.
Obviously, the low mobility value is due to the fact that electron transport occurs through one-dimensional chains, and these channels are easily blocked even by a weak scattering center.
However, for the same reason, these tubes may be promising as sensitive gas sensors.
For such applications, tubes with an odd number of channels seem more attractive because they have a more stable temperature behavior of mobility.








\newpage
\bibliographystyle{ieeetr}
\bibliography{reflist.bib}

\newpage

\end{document}